\documentclass{sf2a-conf}
\usepackage{graphicx}

\begin{document}
\TitreGlobal{SF2A 2007}
\title{Atmospheric dynamics of red supergiant stars and applications to Interferometry. }


\author{Chiavassa A.}\address{GRAAL, cc072, Universit\'e Montpellier II, F-34095 Montpellier cedex 05, France}
\author{B. Plez$^1$} 
\author{E. Josselin$^1$} 
\author{B. Freytag}\address{CRAL, Ecole Normale SupŽrieure de Lyon, F-69364 Lyon cedex 07, France} 

\runningtitle{Atmospheric dynamics of red supergiant stars and applications to Interferometry}

\setcounter{page}{1}

\index{Chiavassa A.}

\maketitle

\begin{abstract}

We have written a 3D radiative transfer code that computes emerging spectra and intensity maps. We derive from radiative hydrodynamic (RHD) simulations of RSG stars carried out with CO$^5$BOLD (Freytag et al. 2002) observables expected for red supergiant stars (RSG) especially for interferometric observations, with emphasis on small scale structures. We show that the convection-related surface structures are detectable in the H band with today's interferometers and that the diameter measurement should not be too dependent on the adopted model. The simulations are a great improvement over parametric models for the interpretation of interferometric observations.

\end{abstract}
%
\section{Introduction}

Red supergiant stars (hereafter RSG) constitute a key-phase in the evolution of massive ($10\leq M_{\rm init}\leq 30 M_{\odot}$) stars. RSG are characterized by a strong mass loss ($M\sim10^{-6}M_\odot$
yr$^{-1}$; Castor 1993) of unknown origin. Furthermore, since RSG are also undergoing a rich nucleosynthesis, they play an important role in the chemical evolution of galaxies.\\
RSGs are irregular, small-amplitude variables and they have a very complex spectrum where strong molecular absorption by TiO and other molecules dominates. This leads to an ill-defined continuum. 
Josselin \& Plez (2007) found from high-resolution spectroscopic time-series observations of a sample of RSGs that the variable velocity fields in their atmosphere probably has a convective origin. The velocities are supersonic and vary with a time-scale of a few $100$ days and the resulting spectral lines have a strong broadening together with asymmetries and wavelength shifts. The observed strong line asymmetries indicate that convection consists of giant cells, as suggested by 3D
radiative hydrodynamic simulations of these stars, performed by Freytag et al. (2002). These models show a peculiar convective pattern, with giant cells evolving on a timescale of about one month with supersonic velocities and shocks. \\
We performed 3D pure LTE radiative transfer calculations in snapshots of these 3D hydrodynamical simulations and we present here the results obtained in terms of spectral synthesis and intensity maps used to create observables for the interpretation of interferometric observations.

\section{Radiative-hydrodynamic simulations}
The numerical simulations analyzed here are performed with
``CO5BOLD'' (``COnservative COde for the COmputation of COmpressible COnvection
in a BOx of L Dimensions, l=2,3'') developed by Freytag, Steffen, Ludwig and collaborators (see Freytag et al. 2002).
These RSG simulations employ the global \emph{star-in-a-box} setup:
the computational domain is a cube, and the grid is equidistant in all directions. Radiation transport is strictly LTE. The grey Rosseland mean opacity is a function of pressure and temperature. The necessary values are found by interpolation in a 2D table which
has been merged at around 12\,000\,K from
high-temperature OPAL data (Iglesias et al. 1992) and low-temperature PHOENIX data (Hauschildt et al. 1997) by Hans-G{\"u}nter Ludwig. Some more technical informations can be found in the CO5BOLD Online User Manual and in the poster contribution by Freytag (this volume).

\section{3D radiative transfer in the hydrodynamical simulations} \label{3dcode}

\begin{figure*}
\centering
\includegraphics[angle=0,width=0.65\hsize]{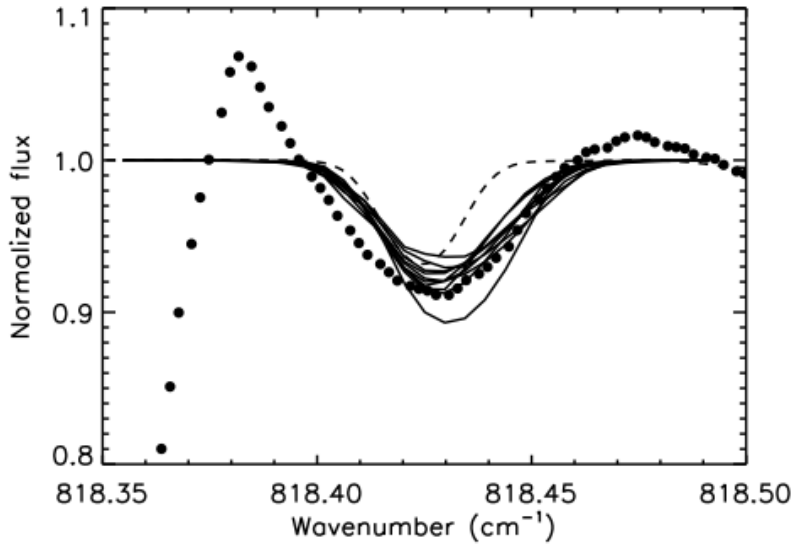} 
\caption{The time sequence computed for the RHD model described in the text (time-step of about $23$ days) is compared to  observations (dots; Ryde et al. 2006; the sharp decrease below $818.38$cm$^{-1}$ is an instrumental artifact). The dashed line is a one-dimensional hydrostatic MARCS (Plez et al. 1992, Gustafsson et al. 2003) model calculation convolved with an exponential profile of $13$km/s width. The line broadening and shifts for the RHD calculations are a natural results of the velocity field of the simulation.}
\label{spec}
\end{figure*}

We performed 3D, detailed, pure LTE radiative transfer calculations in snapshots of the 3D hydrodynamical simulations (the RHD model presented in this work, st35gm03n07, has a solar composition, $235^3$ grid points, M = $12M_{\rm \odot}$, L = $100000L_{\rm \odot}$, T$_{\rm eff}$ = $3482$K, $\log \rm g$ = $-0.39$, and R = $880R_{\rm \odot}$), taking into account the Doppler shifts caused by convective motions. The emerging intensity at the top of one ray of the RHD computational box is computed by summing the contribution of the source function at different depths.\\
To reduce the computing time, the extinction coefficients per unit mass are pre-tabulated as a function of temperature, density and wavelength with a resolution $R=100000$ in the optical range and $R=500000$ in the H and K band. We checked that this resolution is sufficient to ensure an accurate characterization of line profiles even after interpolation at the Doppler shifted wavelengths (Chiavassa et al. 2006).\\
The a posteriori radiation transfer treatment is used to compute: (i) spectra at high and low (Opacity Sampling) spectral resolution that are compared with spectroscopic and spectrophotometric observations; (ii) intensity maps to be compared with interferometric observations.\\
One example is shown on Fig. \ref{spec} where RHD simulations approximately reproduce the $12\mu m$ H$_2$O line width and depth without the need for micro- or macrotubulence  (which is needed in the one-dimensional static models). The shifts of the calculated line center span values between -1 and -3 km/s with respect to the hydrostatic model. The profiles are asymmetric. More observations are needed to check the level of variability.

\section{Intensity maps in the H band}

\begin{figure*}
\centering
\includegraphics[angle=0,width=0.65\hsize]{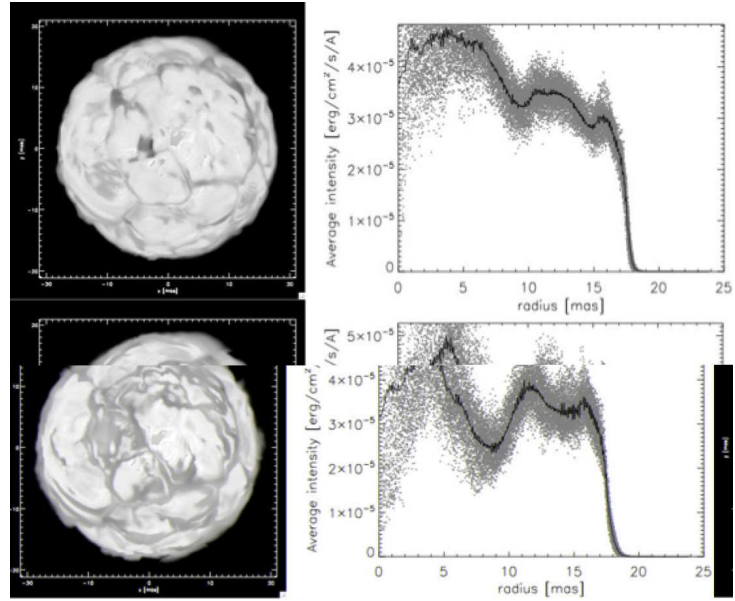} 
\caption{Logarithmic maps of monochromatic intensity (a grey scale is used and lighter shades correspond to higher intensities) and scatter-plot of the radially averaged intensity computed for $48$ snapshots from the RHD simulation described in the text with a time-step of about $23$ days (grey; the black line is the averaged profile). \emph{Upper row}: continuum at $1.6\mu m$ (H$^-$ minimum opacity). \emph{Bottom row}: Nearby CO line.}
\label{intensity}
\end{figure*}

We used our 3D radiative transfer code to derive interferometric observables expected for RSGs with emphasis to small scale structures. We produced a time-sequence, with a time-step of about $23$ days covering about three stellar years, of monochromatic intensity maps at two different wavelengths. These wavelengths probe different depths in the stellar atmosphere since they correspond to the continuum around $1.6\mu m$, where the H$^-$ minimum opacity occurs, and the photosphere becomes more transparent, and a nearby CO line. Two example maps are displayed in Fig. \ref{intensity}, together with the radially averaged intensity distribution. Absolute model dimensions have been scaled to approximately match on angular diameter of $35$mas. The contrast between large bright granules visible in the continuum (upper-left panel), and the smaller structures seen in the CO line (bottom-left panel) is striking. This is due to the convection-related surface structures.

\begin{figure*}
\centering
\includegraphics[angle=0,width=0.65\hsize]{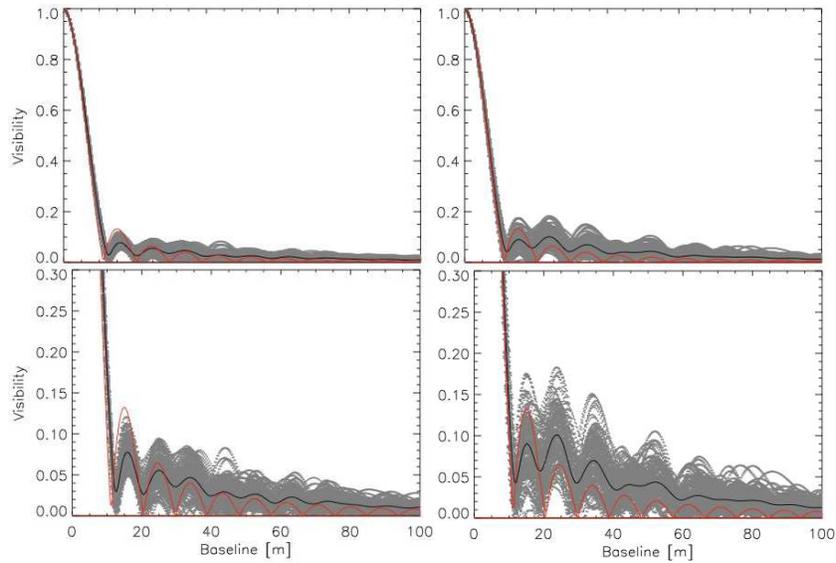} 
\caption{Scatter-plot of visibilities (grey) as a function of projected baseline length, and the average profile (black), for $48$ snapshots and rotated images computed from the RHD simulation described in the text. The red curve is the visibility of a uniform disk with a diameter of $35$mas.  \emph{Left column}:  continuum at $1.6\mu m$. \emph{Right column}: Nearby CO line. Bottom panels are enlargements of the upper ones.}
\label{visibility}
\end{figure*}

\begin{figure*}
\centering
\includegraphics[angle=0,width=1\hsize]{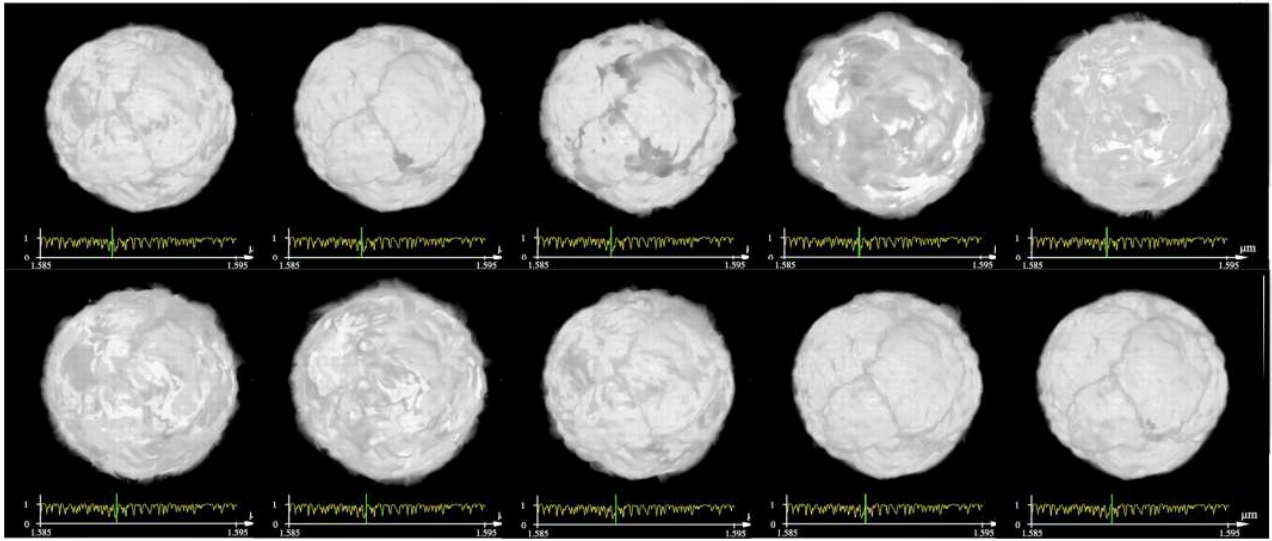} 
\caption{Sequence of intensity maps probing different wavelengths (green vertical line) in the spectrum (yellow line) around the H$^-$ minimum opacity. High spectral resolution in crucial to  determine the variations between different spectral features and between features and continuum (see text).}
\label{seq_carte}
\end{figure*}

We then computed visibility curves from these intensity maps from $48$ snapshots, and also rotating the images (this analysis is extendable to other models and wavelengths). The visibilities carry a clear signature of deviation from circular symmetry (Fig. \ref{visibility}). The small scale structures lead to significant scatter which is larger in the CO line. The first lobe is little affected and thus angular diameter measurement should not be to dependent on the adopted model. We show below that adopting a limb darkened disk model or a uniform disk model results in a $\sim3$\% smaller diameter than using our simulations.\\
The intensity maps in Fig. \ref{seq_carte} probe different wavelengths in the spectrum around the H$^-$ minimum opacity and they depend on the spectral features. Wavelengths probing the continuum show large scale granules; while for wavelengths corresponding to the line, small scale structures appear. Thus, in order to characterize the convective pattern, it is crucial to have the high spectral resolution to determine the variations of the interferometric data between different spectral features, and between features and continuum.\\

The convection-related surface structures are observable with today's interferometers in the H band.

\section{Intensity maps in the K band}

\begin{figure*}
\centering
\includegraphics[angle=0,width=1\hsize]{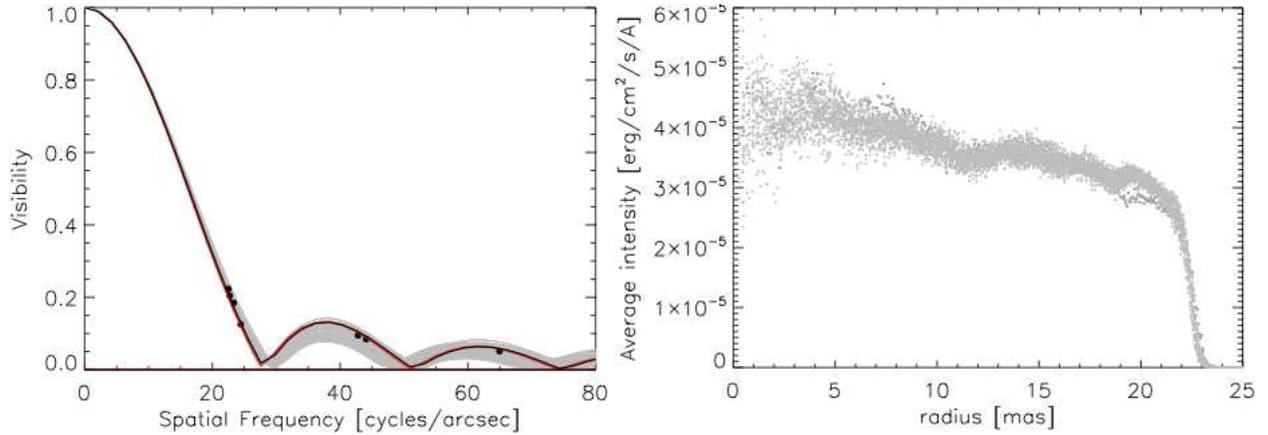} 
\caption{{Left panel: } Synthetic visibilities (grey) compared to $\alpha$ Ori observations in the K band (dots; Perrin et al. 2004a). The red curve is a uniform disk model with a diameter of $43.65$ mas and the black curve (superposed to the red curve) is a limb-darkened disk constructed with the law ($I\left(\mu\right)=1-A\left(1-\mu\right)$) with $A=0.09$ and with a diameter of $43.65$ mas (Perrin et al. 2004). {Right panel: } scatter-plot of the radially averaged intensity computed for $12$ snapshots from the RHD simulations described in the text (grey). The diameter measured by simple models is  $\sim3\%$ smaller than what we find with our calculations.}
\label{perrin1}
\end{figure*}

We employed our code to compute intensity maps also in the K band and we compared our synthetic visibilities with observations of $\alpha$ Ori by Perrin  et al. (2004a). We considered the K222 filter (central wavelength is $2.2$ $\mu m$) mounted on FLUOR (Fiber Linked Unit for Optical Recombination) at the IOTA telescope. Absolute model dimensions have been scaled to fit the observed visibilities. Fig. \ref{perrin1} shows that the observation points are within the scatter due to the small scale convective structures. As for the H band, the synthetic visibilities show the departure from circular symmetry (clearly noticiable in the second and the third lobe in Fig. \ref{perrin1}).\\
Furthermore, we determine the radius of our model from the average intensity profile and we obtained $45$ mas. Below $25$ cycles/arcsec (ie., the first lobe), the convective dispersion seems to be less important and the difference between the diameter measured with the uniform disk model or the limb-darkened disk model and what we found with our calculations is $\sim3\%$. Thus, the diameter measurement should not be too dependent of the chosen model. Nevertheless, from the second to the fourth lobe the information carried by the visibility curves is important to constrain the convection-related surface structures.

\section{Conclusion}

The detection of convection-related granulation on the surface of RSG stars is now possible using our RHD models for the interpretation of the observations. RHD simulations are necessary for a proper qualitative, and quantitative analysis of the surface of cool stars, in order to have parameter-free (without using Mixing-Length theory) estimates of convection. RHD models are \emph{ab initio}, time-dependent, multi-dimensional, non-local and they contain as much physics as possible. The actual limitation is the CPU-time. RHD models are a great improvement over parametric models for the interpretation of interferometric observations, but due to the difficulty of the interferometry, they have to be validated with other techniques such as spectrometry and spectrophotometry (Chiavassa et al. 2006). \\
We found that convection-related surface structures are indeed observable and the interferometric visibilities carry an evident signature of deviation from circular symmetry in the H and K band. The H band is crucial for the detection of the convective signature because small scale structures appear more clearly and contrasted than in the K band. In addition, the first lobe of the synthetic visibilities is little affected by convection, and thus the diameter measurement should not be too dependent on the adopted model; the second, the third and the fourth lobes carry important information because the small wiggle structures visible in the intensity maps (Fig. \ref{intensity}) are not spatially resolved. This is important for the interpretation of forthcoming interferometric observations at high spatial frequencies: in this case, a clear deviation from uniform disk or limb darkened disk models is a signature of the photospheric convection.\\
The analysis shown in this work will be used to set a range for visibility/phase variations due to the photospheric convection to predict further observations and to determine limb-darkening laws for RSGs (Chiavassa et al., in preparation).\\
The work presented here is the first comparison to real interferometric observations. Up to now, there are few available data suitable to constrain the models in terms of visibility and phase. Nevertheless, since RSG are prime targets for Interferometry, thanks to their large diameter and to their high infrared luminosity, a lot of observations should be available in the near future. In this context, we think that a theoretical and observational simultaneous approach is really important to understand the atmospheric dynamics.

\begin{acknowledgements}
We thank the "Centre Informatique National de l'Enseignement Sup\'erieur" (CINES) for 
providing us with computational resources required for part of this work. 
\end{acknowledgements}


\begin{thebibliography}{99}

\bibitem{}Castor, J.~I.\ 1993, ASP 
Conf.~Ser.~ 35: Massive Stars:  Their Lives in the Interstellar Medium, 35, 
297 

\bibitem{}Chiavassa, A., Plez, 
B., Josselin, E., \& Freytag, B., 2006, SF2A-2006: Semaine de 
l'Astrophysique Francaise, 455 

\bibitem{}Freytag, B., Steffen, M., Dorch, B., 2002, Astronomische Nachrichten, 213, 219

\bibitem{}Gustafsson, B., 
Edvardsson, B., Eriksson, K., Mizuno-Wiedner, M., J{\o}rgensen, U.~G., \& 
Plez, B., 2003, Stellar Atmosphere Modeling, 288, 331 

\bibitem{}Hauschildt, Peter H., Baron, E., Allard, France, 1997, ApJ, 390

\bibitem{}Iglesias, Carlos A., Rogers, Forrest J., Wilson, Brian G., 1992, ApJ, 717, 728

\bibitem{}Josselin, E., \& Plez, B.\ 2007, A\&A, 469, 671 

\bibitem{}Perrin, G., Ridgway, S. T., CoudŽ du Foresto, V., Mennesson, B., Traub, W. A., Lacasse, M. G.,  2004a, A\&A, 675, 685


\bibitem{}Plez, B., Brett, J.~M., \& 
Nordlund, A.\ 1992, A\&A, 256, 551 

\bibitem{}Ryde, N., Harper, G.~M., 
Richter, M.~J., Greathouse, T.~K., \& Lacy, J.~H.\ 2006, ApJ, 637, 1040 

\end{thebibliography}
\end{document}